# Electric-field-resolved detection of localized surface plasmons at petahertz-scale frequencies

Dmitry A. Zimin[1,*], Ihor Cherniukh[2,3], Simon C. Böhme[2,3], Gabriele Rainò[2,3], Maksym V. Kovalenko[2,3], Hans Jakob Wörner[1,#]

[1]Laboratory of Physical Chemistry, Department of Chemistry and Applied Biosciences, ETH Zurich, 8093 Zürich, Switzerland

[2]Institute of Inorganic Chemistry, Department of Chemistry and Applied Biosciences, ETH Zürich, 8093 Zürich, Switzerland

[3]Laboratory for Thin Films and Photovoltaics, Empa−Swiss Federal Laboratories for Materials Science and Technology, 8600 Dübendorf, Switzerland

We present a novel electric-field-resolved approach for probing ultrafast dynamics of localized surface plasmons in metallic nanoparticles. The electric field of the broadband carrier-envelope-phase stable few-cycle light pulse employed in the experiment provides access to time-domain signatures of plasmonic dynamics that are imprinted on the pulse waveform. The simultaneous access to absolute spectral amplitudes and phases of the interacting light allows us obtaining a complex spectral response associated with localized surface plasmons. We benchmark our findings against the absorbance spectrum obtained with a spectrometer as well as the extinction cross-section modeled by a classical Mie scattering theory.

**Introduction**

From the perspective of classical physics, a plasmon is a collective oscillation of conduction quasi-free electrons in metal. In quantum physics the plasmon can be described as the quantum of plasma oscillations. In metallic nanoparticles (NPs), the plasmonic oscillations are spatially confined at the interface between a conducting material and the dielectric medium to the length (often defined as the plasmon skin depth) smaller than the wavelength of the driving light and are termed localized surface plasmon (LSP) [1, 2]. Due to the spatial confinement, the plasmonic signatures of metallic NPs differ significantly from those in the respective bulk material [3]. The shape and size of NPs impose additional quantum constraints on the plasmonic dynamics [4]. Common NP materials are noble metals such as silver (Ag) and gold (Au), which exhibit LSP in the visible spectrum [5].

The underlying LSP dynamics could potentially be some of the fastest in optics: their shortest progression times are defined by the inverse spectral width of the plasmonic resonances and are on the order of 100 as (1 as = $10^{-18}$ seconds) [6, 7]. The relaxation times of LSPs are also ultrashort [8-10]. The lifetime is expected to be in the range of 10 – 60 fs (1 fs = $10^{-15}$ seconds) for silver and 1– 10 fs for gold [6]. These ultrafast processes determine the optical properties of plasmonic nanostructures such as resonance frequencies, absorption and scattering cross-sections as well as local-field enhancement factors [3].

Although the understanding of these ultrafast dynamics is far from complete, it is well known that they depend on the composition, size, shape, dielectric environment and, particle–particle separation in NP assemblies. Due to the high sensitivity of LSP responses to the various internal and external degrees of freedom, Ag and Au NPs are extensively used in many applications including surface enhanced spectroscopies [11-13], molecular sensing [14, 15] and photocatalysis

*Corresponding author. Email: d.a.zimin@gmail.com
#Email: hwoerner@ethz.ch

[16-19]. Gold is preferred for biological applications due to its inert nature and biocompatibility [15]. Optical components based on metallic nanostructures have lately emerged as a new promising platform for the manipulation of light [20-22] as well as optical [23-25] and photovoltaic [26, 27] devices. It was also demonstrated that ultrafast plasmonic dynamics can be potentially used for optical memory [28] and information processing at PHz (1 PHz = $10^{15}$ hertz) frequencies [29-34]. For example, ultra-efficient and fast modulators are currently exploited [35] to cope with the increasing demand for higher optical bandwidth for optical communication.

Owing to the manifold of applications, the understanding of ultrafast plasmonic dynamics is of fundamental value. Recent studies have addressed plasmonic temporal [36-44] and spatial [45, 46] dynamics employing conventional spectroscopy, microscopy, interferometry, and pump-probe methods. These methods, however, provide only limited access to the ultrafast plasmonic dynamics as the typical observables are time-integrated spectral intensities or relative spectral phases, with characteristic temporal resolution on the order of tens of fs. The first consequence of such time-integrated observation is its inability to directly access the electric field of oscillating plasmons which inevitably leads to the loss of the information. The second consequence is that the temporal resolution is limited by the temporal envelopes of employed optical pulses. Considering that the durations of typical light pulses in the plasmonic frequency range are on the order of tens of fs, accessing the sub-10 fs dynamics is challenging.

An alternative approach is the electric-field-resolved time-domain spectroscopy [47-49]. This metrology provides full and direct access to the electric field of probing pulses (preserves full temporal information) with a temporal resolution that is not limited by the durations of the employed pulses. The concept was applied to observe the electric field generated by plasmonic nanostructures in the THz (1 THz = $10^{12}$ hertz) domain directly [50]. Unfortunately, the same concept could not be applied to access LSPs dynamics with characteristic resonances at near-PHz frequencies.

Theoretically, the LSP dynamics is commonly addressed by models based on classical Mie scattering, Drude-Lorentz or three-dimensional finite-difference time-domain (FDTD) theories. Mie theory [51] allows the modeling of scattering, absorption, and extinction cross-sections of light interacting with small NPs based on the known dielectric permittivity of a metal and the surrounding medium. It is, however, known that the complex permittivity differs significantly when the NPs dimensions are smaller than the mean free path of the conduction electrons [52, 53]. The Drude-Lorentz approach, on the other hand, models the charge dynamics and associated complex permittivity of a medium. The resulting complex permittivity is a consequence of multiple interplaying mechanisms of charge dynamics that add up together and are represented by phenomenological parameters. Recently, more advanced modified Drude-Lorentz models were proposed in which the bulk and inner electronic levels as well as scattering including electron-electron, electron-phonon, electron-surface and radiation are considered [54 - 56].

Experimentally, the interaction of light with plasmonic NPs results in changes in spectral amplitudes and phases of the interacting light. These changes are a consequence of the light-matter interaction that can be characterized by the complex permittivity of a medium. On the other hand, the complex permittivity is a consequence of the electronic dynamics. Therefore, a simultaneous access to the electric field, or to spectral amplitudes and phases of the interacting light, is necessary for a complete description of the underlying electronic processes. For LSPs in metallic NPs, this, however, requires full access to the electric field of the interacting light at PHz-scale frequencies. In conventional spectroscopy, one typically gets access to only a fraction of the information such as, for instance, an absorbance spectrum derived from the measured light intensities. This loss of information inevitably imposes constraints on the obtainable understanding and modeling of the underlying physics.

Although the measurement of the electric fields of light pulses is an established metrology in the THz or mid-infrared range, the LSP resonances at near-PHz frequencies are challenging to access. Recently, the boundaries of the field-resolved metrology have been extended to the PHz domain [57, 58], providing tools [33, 59-63] that could potentially be employed to directly access the sub-optical-cycle dynamics [64-67] of LSPs with sub-fs temporal resolution.

In this work, we show that a recently demonstrated generalized heterodyne optical-sampling technique (GHOST) [61] can be used for obtaining the full optical response (spectral amplitudes and phases) of LSP dynamics in gold NPs. Our work presents an alternative and simple approach for measuring plasmonic dynamics signatures of which are directly imprinted on the electric field of the interacting light. The direct access paves the way towards accurate modeling of ultrafast processes associated with electronic processes in metallic NPs. By employing a classical Mie scattering theory, we confirm that the measured experimental observables are the signatures of LSPs in gold NPs. The electric field information with sub-fs temporal resolution demonstrated in this experiment might potentially also provide access to the plasmonic build-up and relaxation dynamics that take places within the optical pulse and are not accessible by time-integrated methods.

**Results**

To access the LSPs, we have synthesized spherical polystyrene-capped gold NPs [70] of 20.9 ± 1.2 nm diameter and 30.5 ± 1.7 nm (center to center) interparticle distance. The particles were characterized via transmission electron microscopy. NP thin films have been obtained via spin coating on a thin (~ 170 um) cover glass substrate (more details in the supplementary material).

To preserve the temporal dynamics imprinted on the modification of the test pulse waveform, the plasmonic NPs were deposited at the back surface of the substrate to prevent the dispersion of plasmonic response due to the propagation through the substrate. The substrates with NPs and without NPs were exposed to a few-cycle carrier-envelope-phase (CEP) stable test pulse (~ 5.4 fs, 450 – 1000 nm). After the interaction with the sample, the transmitted test pulse was recombined with the sampling pulse. The recombined sampling and test pulses were guided to the electric-field sampling setup [61]. The recorded GHOST traces of the test pulse with and without NPs are shown in Figure 1B. We observe a clear difference between the recorded waveforms as well as a significant pulse reshaping. To confirm that the measured changes in the pulse waveform are associated with LSPs in gold NPs we analyze the recorded "no NPs" and "with NPs" cases in the spectral domain by applying a Fourier transformation (Figure 2). The transformation allows us to obtain absolute spectral amplitudes and phases across the entire experimental bandwidth (0.3 - 0.65 PHz). We note that the LSP resonance of our gold NPs is around (more details in the supplementary material) 0.53 PHz (570 nm) and therefore well within our experimental range. The results presented in Figure 2 show clear modifications of spectral amplitudes and phases of recorded traces. Interestingly, when comparing spectral amplitudes (Figure 2A) we see a clear reshaping of the pulse spectrum concentrated within about 0.5 – 0.6 PHz range. We observe a similar trend in spectral phases where the largest difference is observed in the same spectral range (0.5 – 0.6 PHz). To confirm that the measured reshaping of the pulse spectrum is a signature of the LSP dynamics, we evaluate the experimental absorbance as $A(f) = \frac{I_{no\,NPs}(f)}{I_{with\,NPs}(f)}$. Here, $I_{no\,NPs}(f)$ and $I_{with\,NPs}(f)$ are spectral intensities of "no NPs" and "with NPs" cases. Figure 3A shows the absorbance resonance centered at about 0.53 PHz. This agrees well with the LSP static absorbance spectra obtained with a spectrometer.

Apart from the absorbance, the signature of the LPS dynamics must be imprinted on spectral phases of the interacting light as well. This information, although not accessible with conventional spectroscopy is crucial for obtaining a complete optical response.

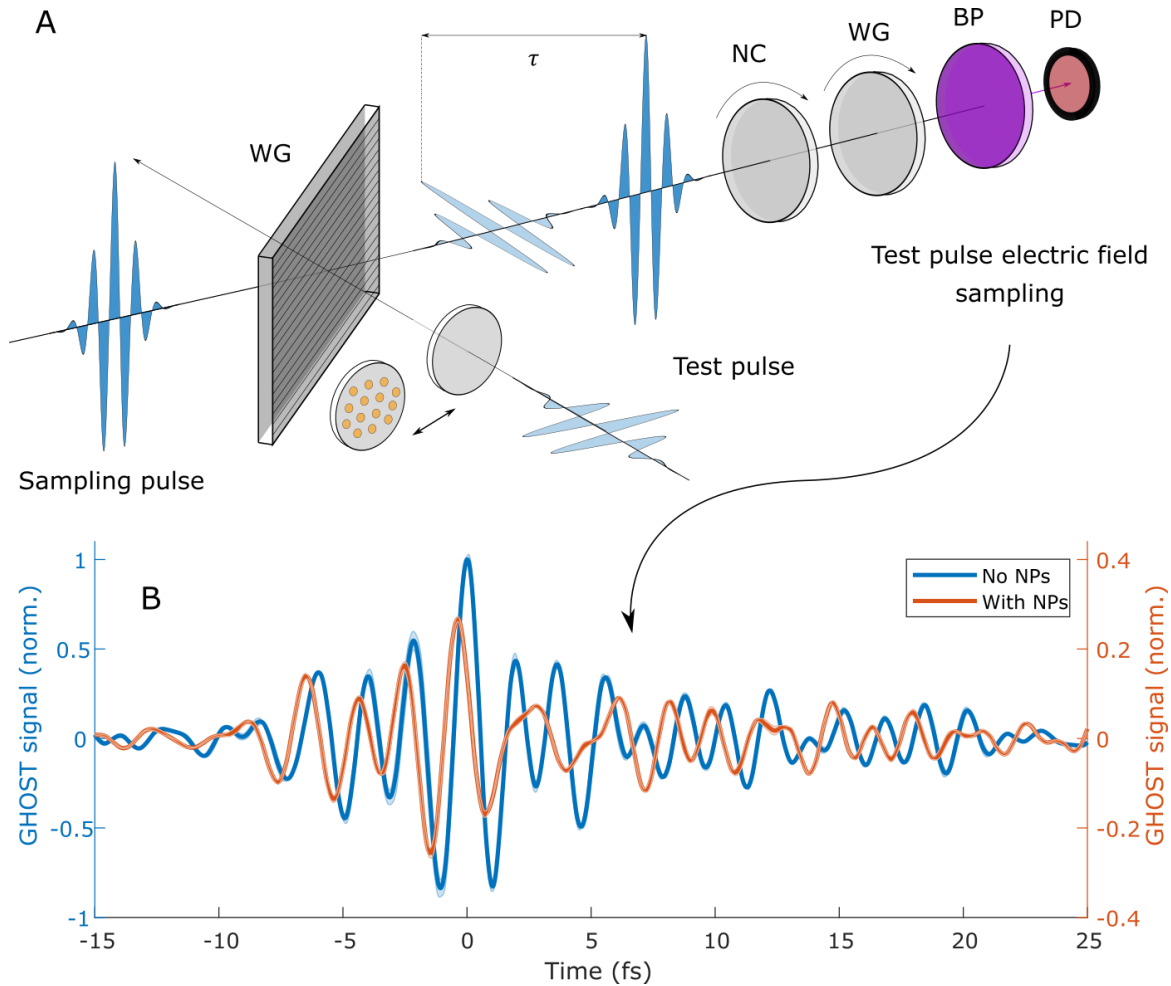

**Figure 1. Experimental concept and results**. (**A**) Schematic of the experimental setup. Two identical cross-polarized pulses named *sampling pulse* and *test pulse* are used in the experiment. The weak test pulse is transmitted through the substrate only and through the substrate with deposited gold NPs. After the light-matter interaction, the transmitted test pulse is recombined with the strong sampling pulse by means of the wire-grid polarizer (WG). Following the recombination, both pulses are incident on a thin z-cut α-quartz crystal (NC) with a controlled time delay $\tau$. The WG after the quartz crystal in combination with the bandpass filter (BP) is used to let the local oscillator and the signal at a frequency of about 0.845 PHz interfere [61]. The intensity of the interference is recorded with the photodiode (PD) at multiple delays $\tau$. The resulting trace is proportional to the electric field of the test pulse [61]. (**B**) Comparison of the measured test pulse GHOST traces after interaction with the substrate only (No NPs; blue trace) and with the substrate coated with gold NPs (With NPs; orange trace). The shadow area around the solid traces corresponds to the standard deviation of five consecutive measurements.

To confirm that our field-resolved experiment is sensitive to this otherwise missing information we apply a similar procedure to derive the modification of the spectral phases, i.e., $\varphi(f) = \varphi_{with\,NPs}(f) - \varphi_{no\,NPs}(f)$ to evaluate signatures of the LSP dynamics encoded in absolute spectral phases. Here, $\varphi_{no\,NPs}(f)$ and $\varphi_{with\,NPs}(f)$ are absolute spectral phases of the "no NPs" and "with NPs" cases. Figure 3B displays the extracted phase difference with a characteristic shape coinciding with the absorbance spectrum. Interestingly, we notice that the extracted phase difference in our experiment is always negative. In a pure Lorentzian resonance, however, one would expect a symmetric phase change around the resonance frequency. This could be a signature of the temporal delay and a potential shift of the CEP. In the frequency domain, the CEP results in a shift of spectral phases by the same constant factor. The time delay on the other hand results in a linear (frequency-dependent) phase shift. Our finding demonstrates that the access to absolute spectral phases can potentially be used to establish an absolute timing of the plasmonic dynamics. To confirm this, we evaluate the group delay $T_g(f) = -d\varphi(f)/df$ of the measured phase response, effectively converting a linear phase shift (time delay) to a constant offset in the frequency-dependent group delay. Figure 3B shows that the evaluated group delay indeed contains a negative offset of about 100 attoseconds. However apart from the delay associated with the temporal response of the LSP dynamics, there might also be a delay associated with the linear propagation through a medium with NPs. In our study we have not attempted to model the LSP dynamics and therefore cannot accurately determine the origin of the measured time delay. However, we note that our experiment is sensitive to such delays due to the absolute spectral phase information originating from the field-resolved nature of the experiment.

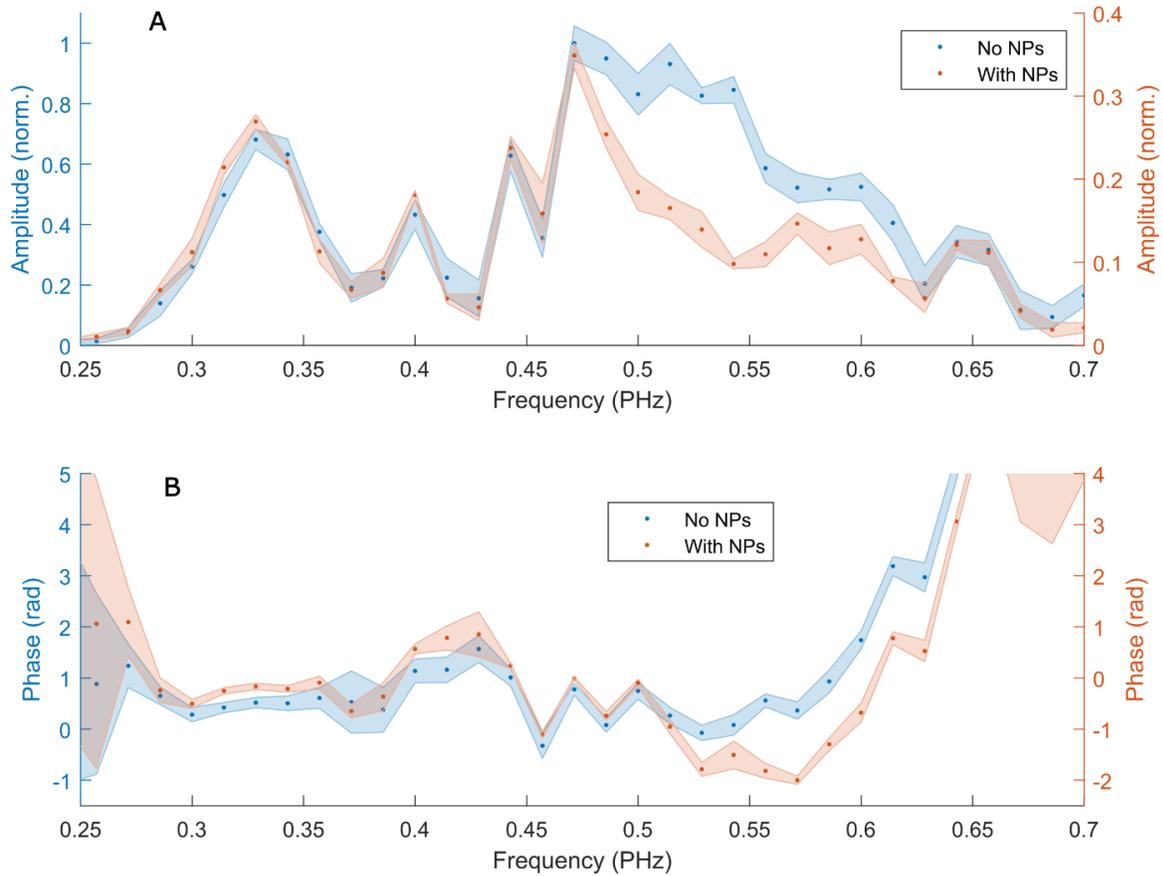

**Figure 2. Frequency domain analysis.** (**A**) Normalized spectral amplitudes of measured reference (no NPs; blue markers) and gold NP thin film (with NPs; orange markers) traces. (**B**) Absolute spectral phases of measured reference and signal traces. The shaded areas correspond to a standard deviation of five consecutive measurements.

Methods for measurement of the electric field of light pulses usually record a signal that is proportional to the electric field. In other words, the recorded signal is a convolution of the electric field $E(t)$ with a response function $R(t)$ of the method in the time domain. In the frequency domain, in terms of spectral amplitudes, the spectral response results in a multiplication of spectral amplitudes $A(f)$ by a frequency dependent spectral response factor $R_A(f)$. For spectral phases $\varphi(f)$, the spectral response results in an addition of the frequency dependent spectral response factor $R_\varphi(f)$. However, since the spectral response in our experiment is the same for "no NPs" and "with NPs" measurements, it is irrelevant when evaluating the absorbance/transmittance and the phase difference between two types of measurements. Therefore, this experimental approach does not require any additional post-processing or reconstruction algorithms.

**Discussion**

After confirming that the field-resolved approach is indeed sensitive to LSPs in gold NPs we notice that our experiment is generally sensitive to the dynamics of not only light absorption but also light-scattering. To disentangle these two contributions, we model our experiment with a classical Mie scattering theory within the frame of the quasi-static approximation. Generally, the theory allows modelling of the polarizability ($\alpha$), extinction ($C_{ext}$), absorption ($C_{abs}$), and scattering ($C_{sca}$), cross-sections of a spherical NP with a known radius ($R$) and the dielectric functions of the metal ($\varepsilon_r$) and the surrounding medium ($\varepsilon_m$):

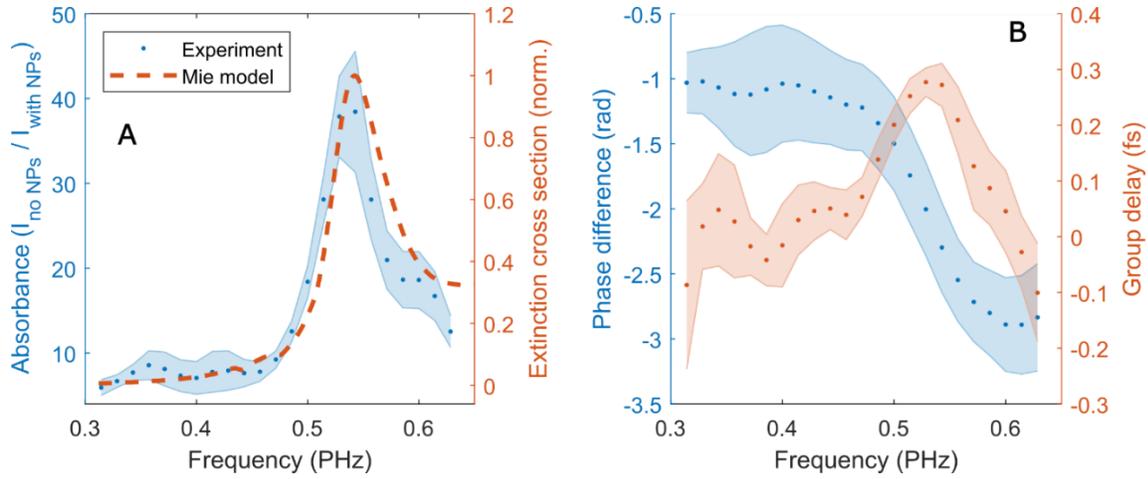

**Figure 3. Spectral response of gold NPs.** (**A**) Experimental absorbance of a gold NP thin films (blue markers) and the modelled extinction cross-section employing Mie scattering theory (red dashed line). (**B**) Experimental spectral phase difference (blue markers) and group delay (orange markers) caused by the gold NP thin film. The shaded areas correspond to a standard deviation of twenty-five combinations of the measured "no NPs" and "With NPs" cases.

$$\alpha(f) = 4\pi R^3 \frac{\varepsilon_r - \varepsilon_m}{\varepsilon_r + 2\varepsilon_m}$$

$$C_{sca}(f) = \frac{k_m^4}{6\pi}|\alpha(f)|^2$$

$$C_{ext}(f) = k_m \Im[\alpha(f)]$$

$$C_{abs}(f) = C_{ext}(f) - C_{sca}(f)$$

Here $\Im$ stands for the imaginary part, while $k_m$ is the wavevector. For modelling we used known complex refractive indices of gold [68] and polystyrene (surrounding medium) [69]. The radius of NP in the model was set to 10.45 nm according to the sample characterization (more details in the supplementary section). The wavevectors $k_m$ were calculated according to the experimental frequencies. Figure 4A demonstrates that due to the small size of NPs, the extinction cross-section is almost entirely coinciding with the absorption cross-section. The scattering cross-section (Figure 4B) is almost two orders of magnitude smaller than that for the absorption. These modelling results suggest that our experimental observations are almost entirely associated with the absorption of light by gold NPs. We note that this conclusion is useful for theoretical description of LSP dynamics where scattering of light can be neglected.

**Conclusion**

We have demonstrated a novel approach to access ultrafast LSP dynamics in metal (Au) NPs at near-PHz frequencies. In contrast to conventional spectroscopy, interferometry and pump-probe methods, our approach yields the complete complex spectral response of LSP dynamics that is imprinted on absolute spectral amplitudes and phases of the interacting light pulse. Due to the field-resolved nature of the detection, this approach also offers sub-fs temporal resolution, hereby surpassing conventional time-integrated methods. The approach does not require additional post processing or reconstruction.

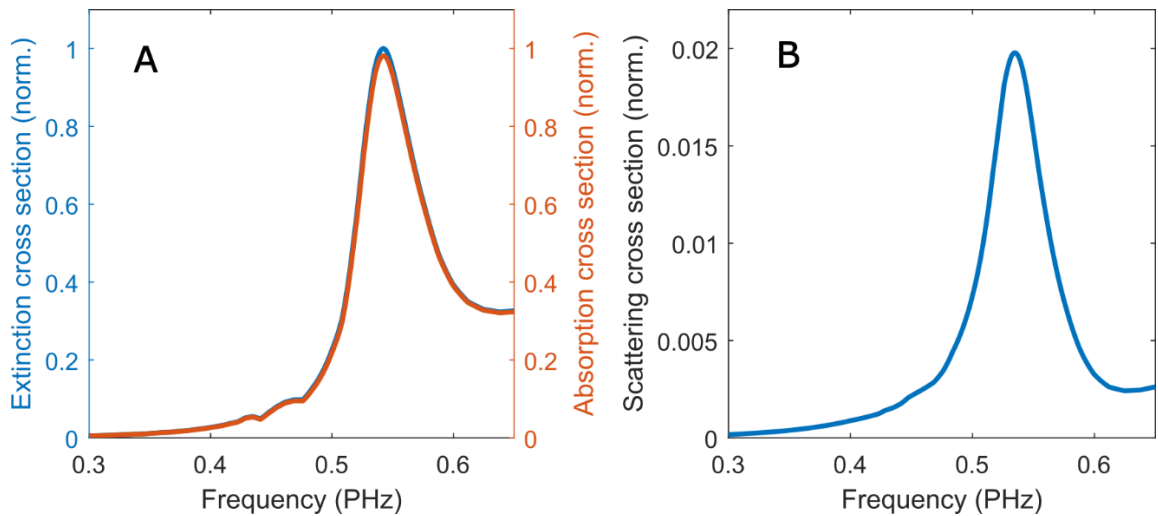

**Figure 4. Extinction, absorption, and scattering cross-sections.** (**A**) Normalized extinction (blue trace) and absorption (red trace) cross-sections of a spherical gold NP of 20.9 nm diameter (as in the experiment). (**B**) Normalized scattering cross-section of a spherical gold NP under experimental conditions.

We observe that the GHOST signal of the test pulse transmitted though the substrate differs significantly when compared to the signal transmitted though the substrate with deposited gold NPs. We confirm that this difference is a fingerprint of LSP dynamics in gold NPs. By modelling the experiment with a classical Mie scattering theory, we find that our experimental observations are dominated by the charge dynamics associated with the absorption of light. The direct access to the complex optical response may pave the way towards accurate modelling of the LSP dynamics that is confined within the interacting optical pulses and therefore not accessible by time-integrated method.

**Materials and Methods**

L-ascorbic acid (≥ 99.9 %, Sigma-Aldrich, CAS 50-81-7); hydrogen tetrachloroaurate hydrate (Acros Organics; CAS 27988-77-8); hexadecyltrimethylammonium chloride (CTAC, 99 %, Thermo Scientific Chemicals); hexadecyltrimethylammonium bromide (CTAB, > 98.0 %, TCI); sodium borohydrate (> 96 %, Sigma-Aldrich); polystyrene w-thiol-terminated (PSS, Mn = 6,500 g mol$^{-1}$, PDI:1.18, Polymer Source); toluene (≥ 99.8 %, Fisher Chemical); tetrahydrofuran (99.9 %, Sigma-Aldrich);

**Acknowledgments:** The authors acknowledge A. Schneider, M. Seiler, M. Urban for technical support with mechanical and electronic parts of the experiment, J. Bredehoeft, D. T. Matselyukh and J. Wiese for construction, characterisation, and operation of the few-cycle laser, D. Schauer for synthesis of test Au-NP targets and J. Ji for helping with the data analysis.

**Funding**: This study is supported by the ETH Zürich (H. J. W.), ERC Grant #772797 (H. J. W.) and ETH Zürich Career Seed Award (D. A. Z.)

**Author contributions:**
Conceptualization: D. A. Z.
Methodology: D. A. Z., S. B., I. C., M. V. K.
Investigation: D. A. Z.
Visualization: D. A. Z., S. B., I. C.
Supervision: H. J. W., M. V. K., D. A. Z.
Writing—original draft: D. A. Z., S. B.
Writing—review & editing: D. A. Z., I. C., S. B., G. R, H. J. W, M. V. K.

**Competing interests:** Authors declare that they have no competing interests.

**Data and materials availability:** All data are available in the main text or the supplementary Materials

# Supplementary Materials for

# Field-resolved detection of localized surface plasmons at petahertz-scale frequencies

Dmitry A. Zimin[1*], Ihor Cherniukh[2], Simon C. Böhme[2], Gabriele Rainò[2], Maksym V. Kovalenko[2], Hans Jakob Wörner[1,#]

*Corresponding author. Email: d.a.zimin@gmail.com
#Email: hwoerner@ethz.ch

## S1 Laser beamline

A Ti:Sa oscillator (Rainbow, Femtolasers GmbH) provides a spectrum centered at ∼ 790 nm. The output from the oscillator is guided through the CEP stabilization module based on the feed-forward scheme. CEP stable pulses are further pre-amplified within a multi-pass Ti:Sa chirped-pulse amplifier at a repetition rate of 1 kHz. The pre-amplified pulses are further amplified in an additional multi-pass chirped-pulse amplifier. After the amplification, the pulses are compressed with a transmission grating based compressor yielding ∼ 25 fs pulses of about 5 W average power. A part (about 1.6 W) of the amplified pulses is spectrally broadened in a hollow-core fiber (HCF), filled with 1.6-bar He gas for spectral broadening. The home-built chirped-mirror compressor consisting of chirped mirrors (PC70 from Ultrafast Innovations) in combination with fused silica glass and air is then used to compress the broadened spectrum to ∼ 5.4 fs pulse duration. A reflection of the glass is guided to the active CEP stabilization setup. The active stabilization is based on the f-to-2f technique.

## S2 Experimental setup

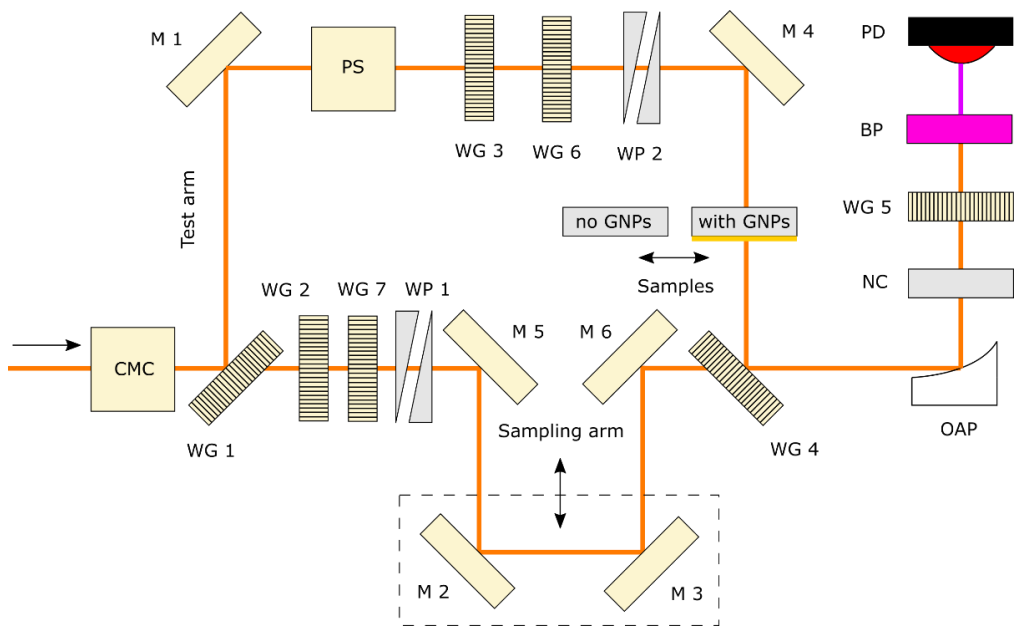

**Figure S1. Experimental setup.** WG – wire grid polarizer, WP – wedge pair, M – protected silver mirror, NC – non-linear crystal, BP – bandpass filter, PD – photodiode, CMC – chirped mirror compressor, PS – periscope.

The experimental setup is shown on Figure S1. After the chirped-mirror compressor (CMC), the pulse is split into two optical arms (sampling and test). The test arm is guided through a periscope consisting of two protected silver mirrors for rotation of the polarization by 90 degrees. After the periscope, the test pulse is transmitted through the set of wire gird polarizers (WG 3, WG 6) for adjusting of the pulse energy. The wedge pair (WP 2) is used for the fine-tuning of the compression of the test pulse. After the setting of the test pulse energy and compression, the pulse is transmitted either through the reference (no GNPs; substrate only) or through the sample (with GNPs; gold NP thin film on substrate) and recombined with the sampling arm by the wire grid polarizer WG 4.

The sampling pulse undergoes a similar procedure as the test pulse where the wire grid polarizers WG 2 and WG 7 are used to set the pulse energy while the wedge pair WP 1 is used for fine-tuning of the pulse compression. The delay line consisting of M 2 and M 3 protected silver mirrors is placed on a linear piezo stage for adjusting of the delay between sampling and test pulses.

After recombination of the sampling and test pulses with the wire grid polarizer WG 4, both cross-polarized pulses are focused with the off-axis protected silver parabolic mirror (OAP) on the non-linear crystal (NC). The non-linear crystal is a z-cut α-quartz. The emerging local oscillator and a signal are chosen [61] by means of the wire grid polarizer WG 5, while the detection frequency is defined by the narrow bandpass filter centered at about 0.845 PHz. The interfering local oscillator and the signal are incident on a GaP-based photodiode for GHOST electric field sampling [61].

## S3 Data acquisition

The experimental signal corresponds to the voltage measured from the photodiode with a lock-in amplifier at each time delay between test and sampling pulses. Instead of blocking every second pulse using a mechanical chopper placed in one of the optical arms we flip the CEP of every second pulse entering the experimental setup by π with the help of the DAZZLER software. We lock the lock-in amplifier at half of the repetition rate (~500 Hz) for the signal detection. In contrast to the mechanical chopper scheme, this method [59] provides us with an increase of the signal-to-noise ratio. The voltage was readout by a software via GPIB interface (National Instruments) and stored on a computer.

## S4 Sample preparation and characterization

Colloidal Au nanoparticles (NPs) with PSS ligands were synthesized employing a two-step synthesis approach and a subsequent ligand exchange to polystyrene following previously published procedures [70, 71]. In a first step, small Au seeds were produced by rapidly injecting aqueous sodium borohydride (600 µL, 0.01 M) into a mixture consisting of CTAB (9.9 mL, 0.1 M) and hydrogen tetrachloroaurate (100 µl, 0.025 M) under vigorous stirring (1000 rpm). Subsequently, the stirring rate was reduced to 300 rpm and maintained for 3 min. The reaction mixture was then left undisturbed for 3 h. Next, 500 µl of the seeds solution was combined with ascorbic acid (15 ml, 0.1 M) and CTAC (20 ml, 0.2 M), and the mixture was stirred at 700 rpm. A one-shot injection of hydrogen tetrachloroaurate (10 ml, 2.2 mM) was then performed. After 15 min of stirring at 300 rpm, the Au NCs were subjected to centrifugation at 12,100 rpm for 60 min at 10 °C. The resulting precipitate (approximately 0.2–0.3 ml of concentrated solution) was subsequently redispersed in 1 ml of 0.02 M CTAC, centrifuged again at 12,100 rpm for 20 min and redispersed in 1.3 ml of 0.02 M CTAC. In the final growth step, the Au NCs were mixed with CTAC (35 ml, 0.1 M). The mixture was then stirred at 400 rpm in a water bath at 30 °C, and ascorbic acid (4.7 ml, 22.6 mM) was added. Approximately one minute later, hydrogen tetracholoroaurate (10 ml, 6 mM) was slowly added at a rate of 6.67 mL h$^{-1}$, using a syringe pump. Once the addition was completed, the mixture was stirred at 400 rpm for another 10 min at 30 °C and then centrifuged at 12,100 rpm for 20 min at 11 °C. The NPs (with about 150 µL of supernatant) were redispersed in 3 ml of water. The solution was centrifuged at 1,500 rpm for 3 min and the precipitate was discarded. The resulting supernatant was centrifuged again at 12,100 rpm for 10 min.

To obtain a colloid which would allow the processing of thin films with suitable inter-NP spacing, we applied a post-synthetic ligand exchange. An aqueous solution of NPs (in ~ 200 µl water) obtained after the last centrifugation was added to a solution of PSS-6.5k (15 mg, in 2 ml tetrahydrofuran) under vigorous and prolonged (overnight) stirring. The product was washed four times with tetrahydrofuran and ethanol as solvent/antisolvent pair (1:1 – 1.5 by volume) and once with toluene and ethanol as solvent/antisolvent pair (1:1 by volume) and dissolved in toluene.

A thin film of Au NPs on a glass substrate was obtained via spin coating 60 µl of the final colloid at 1,200 rpm for 60 s.

Figure S1 shows a TEM image of a monolayer of Au NPs. The NPs are of spheroidal shape, with a mean diameter of 20.9 ± 1.2 nm and a mean interparticle distance of 30.5 ± 1.7 nm (center to center). The error bars denote the standard deviation of the Gauss function fitted to the NP size histogram (upper inset) and inter-NP distance histogram (bottom inset), respectively.

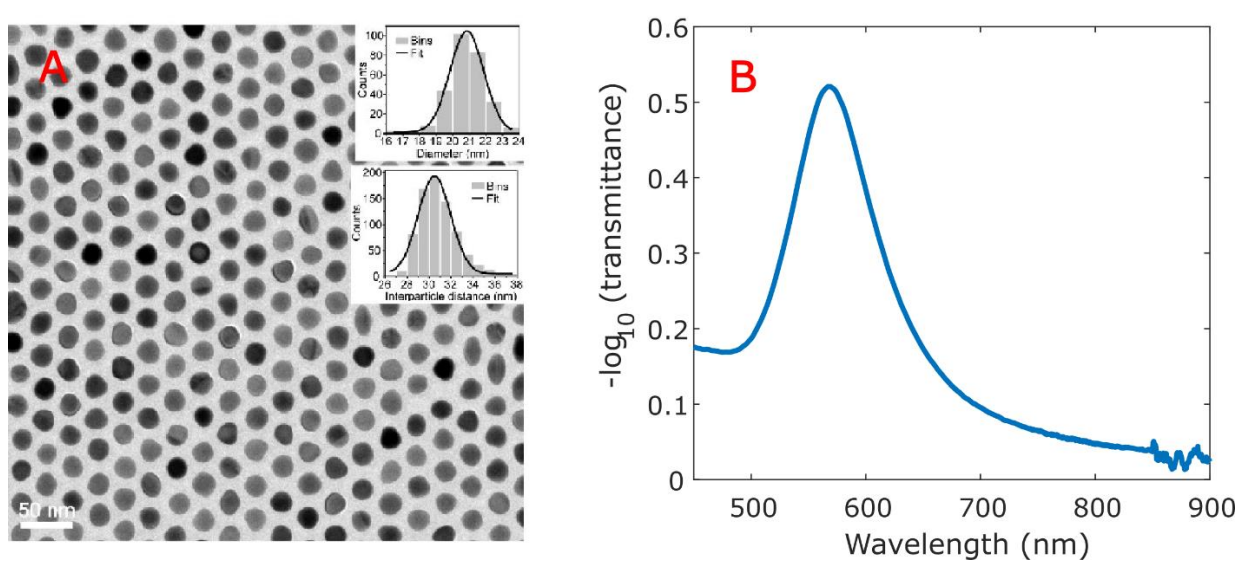

**Figure S2.** (**A**) TEM image of an Au NP monolayer. Upper inset: NP size histogram, returning a mean NP size of 20.9 ± 1.2 nm; bottom inset: inter-NP (center-to-center) distance histogram, returning a mean distance of 30.5 ± 1.7 nm. (**B**) Absorbance spectrum measured with a spectrometer.